# IoMT-Blockchain based Secured Remote Patient Monitoring Framework for Neuro-Stimulation Device

Md Sakib Ullah Sourav[1], Mohammad Sultan Mahmud[2], Md Simul Hasan Talukder[3], Rejwan Bin Sulaiman[4], Abdullah Yasin[5]

[1]School of Management Science and Engineering, Shandong University of Finance and Economics, China
[2]College of Computer Science and Software Engineering, Shenzhen University, China
[3]Bangladesh Atomic Energy Regulatory Authority, Bangladesh
[4]School of Computer Science and Technology, Northumbria University, United Kingdom
[5]Department of Electrical and Electronics Engineering, American International University, Bangladesh

**Emails**: sakibsourav@outlook.com; sultan@szu.edu.cn; simulhasantalukder@gmail.com; rejwan.binsulaiman@gmail.com; yasin1758@gmail.com

*Abstract*— Biomedical Engineering's Internet of Medical Things (IoMT) is helping to improve the accuracy, dependability, and productivity of electronic equipment in the healthcare business. Real-time sensory data from patients may be delivered and subsequently analyzed through rapid development of wearable IoMT devices, such as neuro-stimulation devices with a range of functions. Data from the Internet of Things is gathered, analyzed, and stored in a single location. However, single-point failure, data manipulation, privacy difficulties, and other challenges might arise as a result of centralization. Due to its decentralized nature, blockchain (BC) can alleviate these issues. The viability of establishing a non-invasive remote neurostimulation system employing IoMT-based transcranial Direct Current Stimulation is investigated in this work (tDCS). A hardware-based prototype tDCS device has been developed that can be operated over the internet using an android application. Our suggested framework addresses the problems of IoMTBC-based systems, meets the criteria of real-time remote patient monitoring systems, and incorporates literature best practices in the relevant fields.

## 1. INTRODUCTION

The Internet of Things (IoT), Artificial Intelligence (AI), Machine Learning (ML), Robotics, Blockchain, and other smart technologies have revolutionized engineering and manufacturing. The healthcare industry is no exception. The healthcare system has been boosted by new technology, making it more viable and accessible to the general public. Since the 1970s, there has been a substantial change in the way technology is used in this field. Because of the rapid expansion and use of the Internet of Things (IoT) and Cloud computing, today's generation, known as Industry 4.0, is entirely reliant on intelligent gadgets and their use. IoT is used in a variety of fields, including smart cities, smart homes, smart grids, security and emergency situations, smart agriculture, smart monitoring, and so on [1]. Many sensors are employed for use in the creation of various intelligent wearable devices that aid in the monitoring of human activities as well as the recording of health data. In the field of biomedical engineering, the internet of medical things (IoMT) allows doctors to treat patients remotely. Models must be in place, however, to guarantee that the treatments are carried out properly, considering the security issues connected with the IoMT [2]. COVID-19's recent events demonstrate the need for a remote patient monitoring (RPM) system to be developed because of the benefits of protecting vulnerable patients by reducing the need for visits to hospitals and other clinical settings, as well as reducing risk to physicians by reducing physical patient contact [3].

For many years, neurorehabilitation has employed transcranial direct current stimulation (tDCS) to effectively boost or reduce mental function and learning [4] and it is considered to be safe and widely accepted [5]. Several studies have looked at using transcranial direct current stimulation (tDCS) to treat neurological illnesses including Parkinson's disease and other movement-related disorders [6, 7]. Such therapies have been shown to be successful in a wide spectrum of individuals with neuro diseases, with tDCS therapy improving quality of life (QoL) in people who would otherwise suffer greatly [8]. tDCS systems require researchers to collaborate with patients in order to achieve the desired outcomes. This is due to the fact that these systems must be able to accurately target and focus on certain parts of the brain in order to stimulate them [9]. Furthermore, tDCS devices can be quite costly, limiting their application to specialized units with available resources [10]. As a result, new methods for conducting tDCS have been developed in order to improve patient outreach. For instance, the approaches described by Sourav et al. [11] and Samuel et al. [12] are based on an open source framework and deliver the same medicines to a larger number of patients. However, such systems are still in the early stages of research and are subject to restrictions such as the accuracy of real output currents and the system's effectiveness. Treatment monitoring and delivery are crucial components of any new tDCS system, as tDCS therapies require specialized clinical supervision. More patients might be served by providing remote and cloud-based services for such therapy. This research will look into how clinicians can use tDCS remotely using cloud-based tools. Legal and ethical considerations, as well as the requirement for safe testing and development prior to clinical trials, must all be considered in such a system. As a consequence, these automated remote solutions must be safe and secure, with no risk of patient privacy or tDCS treatment abuse or misuse. This project will look into using off-the-shelf components in the hardware design to keep costs down and make the device more accessible for patients.

Doctors rely on sensitive healthcare data to monitor and diagnose their patients' health. Data acquired from devices or diagnoses of patients is frequently kept in a centralized system called Electronic Health Records (EHR) for later study. Storing all of a patient's data in one location increases the danger of data loss, manipulation, and hacking. Furthermore, storing all data in a central repository makes it impossible to establish transparency. Furthermore, these records should be individualized and accessible to both patients and physicians from any location, which may be accomplished with the use of a decentralized storage system. However, data decentralization must ensure that data is not tampered with and is accessible to everyone in a secure manner. To address all the issues, blockchain, a secured and consistent decentralized storage technology can be introduced.

Nonetheless, integrating blockchain with IoT in healthcare is difficult, and there are just a few researches on the topic. This paper has a two-fold contribution from us.

- We created a hardware prototype of a tDCS device with unique characteristics that may be used at home with real-time guidance and instructions from a doctor.
- To incorporate in our suggested model, we evaluated the current literature that blends IoMT with blockchain.

This article is structured in the following sections. Brain simulation and devices are described in Section 2. Section 3 discusses the optimal conditions reported in the literature for tDCS treatments. Section 4 highlights the specifications of our proposed framework. Finally, Section 5 concludes the work with future directives.

## 2. BRAIN SIMULATION AND DEVICES

Brain stimulation is emerging as a highly promising treatment option for a wide range of disorders, most notably epilepsy. This cutting-edge approach involves the precise application of scheduled stimulation to specific cortical or subcortical targets, facilitated by commercial devices designed to deliver electrical pulses at designated intervals. The primary goal of this technique is to effectively modify the intrinsic neurophysiologic properties of epileptic networks, potentially offering transformative therapeutic outcomes. Among the extensively researched targets for scheduled stimulation, the anterior nucleus of the thalamus and the hippocampus have garnered considerable attention. Studies have demonstrated that activating the anterior nucleus of the thalamus can lead to a significant reduction in seizures, even months after the stimulator implantation [13]. Moreover, exciting progress has been made in treating cluster headaches (CH) through the use of temporary stimulating electrodes at the sphenopalatine ganglion (SPG), with patients reporting rapid pain relief within minutes of the stimulation [14].

The advancement of brain stimulation techniques has not been limited to invasive approaches involving implanted electrodes. Pioneering researchers have taken non-invasive strides by engineering a transparent zirconia "window" implanted in mice skulls. This ingenious method allows optical waves to penetrate more deeply, akin to the principles of optogenetics. By leveraging this non-invasive technique, researchers can precisely stimulate or inhibit individual neurons, broadening the horizons of brain research and fostering potential therapeutic breakthroughs.

### 2.1. INVASIVE OF BRAIN STIMULATION

Invasive techniques involve surgical procedures to implant electrodes or other devices directly into the brain to deliver electrical impulses or stimulation. The invasive stimulation devices are listed in Table 1.

**Table 1**. Invasive neurostimulation devices.

| Serial No | Name of the noninvasive stimulation |
|---|---|
| 1 | Deep Brain Stimulation (DBS) |
| 2 | Epidural Cortical Stimulation |

### 2.1.1. DEEP BRAIN STIMULATION (DBS)

Deep Brain Stimulation (DBS) is a surgical procedure and neuromodulation technique used to treat certain neurological conditions by delivering electrical impulses to specific brain regions [15]. It involves implanting a small, battery-operated medical device, often referred to as a "brain pacemaker," into the brain. The device consists of electrodes that are carefully positioned in targeted brain areas and connected to a pulse generator, typically implanted under the skin in the chest or abdomen [16].

The working principle of DBS revolves around modulating abnormal neural activity in the brain circuits associated with various movement and neuropsychiatric disorders [17]. The electrical stimulation delivered by the electrodes helps regulate the firing patterns of neurons, effectively suppressing or facilitating specific brain pathways, which can alleviate symptoms and improve overall brain function.

### 2.1.2. EPIDURAL CORTICAL STIMULATION

Epidural Cortical Stimulation (ECS) is a relatively novel brain stimulation technique that involves the placement of electrodes on the surface of the brain, specifically on the cerebral cortex, and delivering electrical impulses to modulate brain activity [18]. ECS is a form of neuromodulation and is used to explore brain function, investigate neural circuits, and potentially treat certain neurological and psychiatric disorders [19]. The procedure for ECS typically involves a surgical implantation of a thin sheet or grid of electrodes directly on the surface of the brain, just beneath the dura mater (the outermost protective membrane surrounding the brain). These electrodes can then be used to apply electrical currents to the cortical surface, allowing researchers or clinicians to stimulate or inhibit specific brain regions or neural circuits.

ECS is believed to work by directly influencing the electrical activity of the targeted brain areas [20]. By modulating neural firing patterns and synaptic transmission, ECS can potentially alter brain network dynamics and influence various cognitive and motor functions.

As a research tool, ECS allows scientists to study brain function with high spatial and temporal resolution, providing valuable insights into brain organization and connectivity. In a clinical context, ECS is being investigated as a potential treatment option for conditions like epilepsy, chronic pain, movement disorders, and even certain cases of traumatic brain injury or stroke rehabilitation [21]. The therapeutic potential of ECS is still being explored, and its use in clinical settings is limited to specialized centers and research studies.

It is essential to note that ECS is an invasive procedure and carries potential risks, including infection, bleeding, and damage to brain tissue. As with other brain stimulation techniques, careful patient selection, thorough evaluation, and appropriate post-operative care are crucial to ensure safety and optimize outcomes. Additionally, since the field of neuroscience and neuromodulation is continuously evolving, there might have been further advancements or updates in ECS research beyond my knowledge cutoff date.

### 2.2. NON-INVASIVE OF BRAIN STIMULATION

Non-invasive techniques do not require surgery and involve the application of external stimuli to the scalp or other peripheral areas to influence brain activity. The non-invasive neurostimulation devices are enlisted in Table 2.

**Table 2.** Non-Invasive neurostimulation devices

| Serial No | Name of the non-invasive stimulation |
|:---:|:---:|
| 1 | Transcranial Magnetic Stimulation (TMS) |
| 2 | Transcranial Direct Current Stimulation (tDCS) |
| 3 | Transcranial Alternating Current Stimulation (tACS) |
| 4 | Transcranial Random Noise Stimulation (tRNS) |

### 2.2.1. TRANSCRANIAL MAGNETIC STIMULATION (TMS)

Transcranial Magnetic Stimulation (TMS) is a non-invasive medical procedure used to treat certain neurological and psychiatric conditions [22]. It involves the use of electromagnetic induction to create small electrical currents in specific areas of the brain.

During a TMS session, a magnetic coil is placed against the scalp of the patient. When an electrical current pass through the coil, it generates a magnetic field that can penetrate the skull and stimulate the underlying brain regions [23]. The stimulation can either increase or decrease the activity of the targeted brain area, depending on the frequency and intensity of the magnetic pulses. The structure of it is depicted in Figure 1.

There are two main types of TMS:
- Repetitive Transcranial Magnetic Stimulation (rTMS): In this method, multiple magnetic pulses are delivered in rapid succession to the targeted brain region. rTMS can either increase or decrease neuronal activity and is often used as a therapeutic tool for various neurological and psychiatric disorders.
- Deep Transcranial Magnetic Stimulation (dTMS): dTMS is a variation of TMS that uses H-coils to target deeper brain structures. It is commonly used to treat conditions like depression and obsessive-compulsive disorder.

TMS has shown promise as a treatment option for various conditions, including depression, anxiety disorders, migraines, and certain types of chronic pain. However, its exact mechanisms of action are still not fully understood, and research in this field is ongoing.

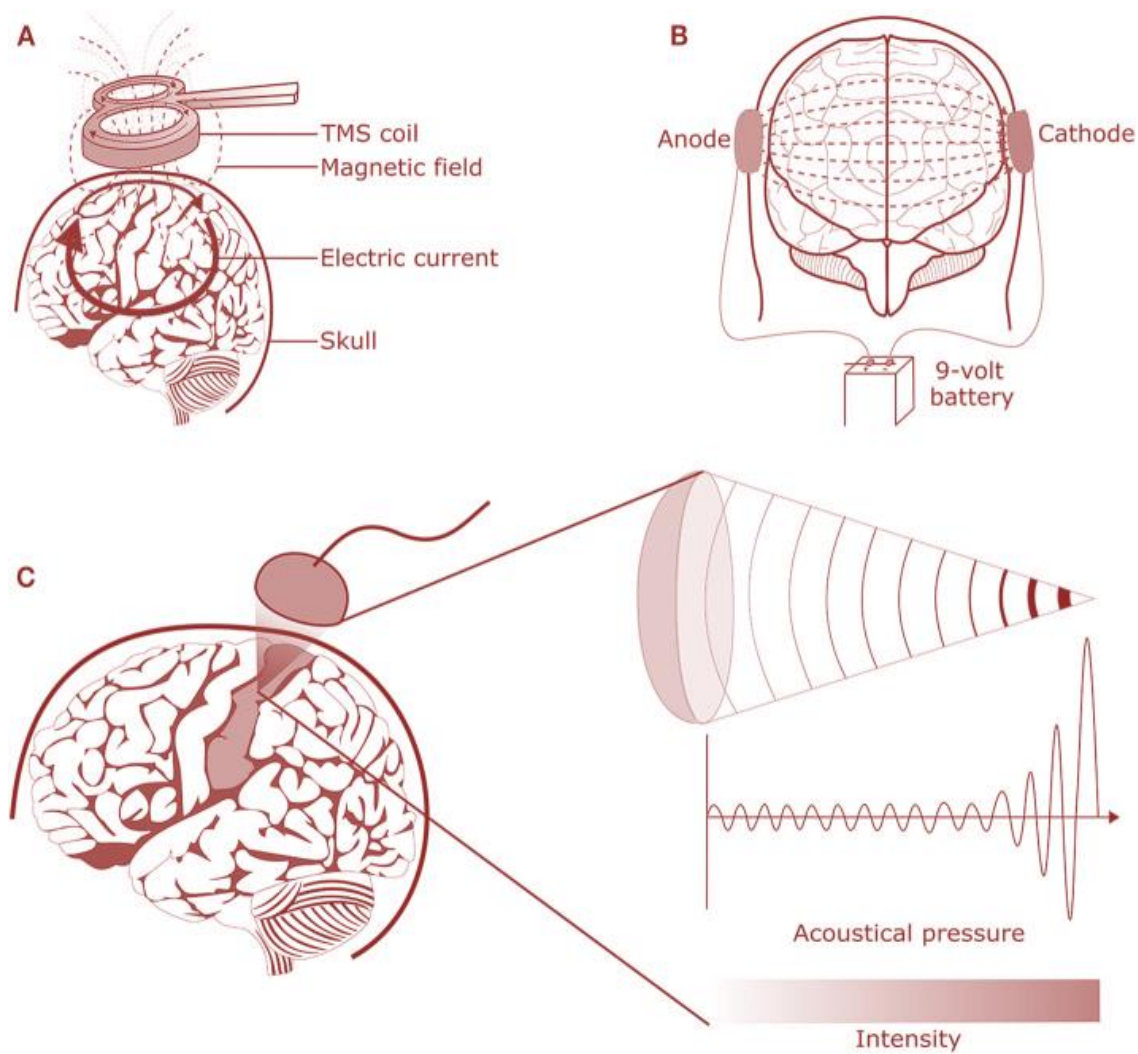

**Figure 1.** TMS of the brain.

### 2.2.2. TRANSCRANIAL DIRECT CURRENT STIMULATION

The transcranial direct current stimulation (tDCS) is a non-invasive brain stimulation technology that is used to control the excitability of the central nervous system in individuals [24]. The goal of central nervous system stimulation is to alter the firing of neurons in the brain. The impact of the changed neurons might be beneficial or harmful to a patient. A lot of studies have looked at the best testing criteria for individuals receiving tDCS therapy. Session lengths (minutes), current dosages (mA), and session timelines are among the factors. The goal is to figure out how to provide the best circumstances for long-term cognitive plasticity enhancement [25].

Bikson et al. [26] defined the safety limits for tDCS treatments, recommending a treatment length of 20 minutes on average, with a range of 5-30 minutes. Thair et al. [27] verified that the duration of therapy is determined by the neuro physician's prescription for each session. As a result, any tDCS system that is built must be capable of operating optimally for the duration of the therapy, which might be up to 30 minutes [26,27]. In addition, studies have been undertaken to assess if current tDCS levels are both safe for patients and give enough stimulation to achieve beneficial effects. According to Parazzini et al. [28], 1 mA has no brainstem interference and is therefore a suitable dosage for sustained tDCS therapy of up to 30 minutes. Parazzini et al. [29] showed a current dosage of less than 2 mA had no effect on the heart, indicating a safe current range of 1 to 2 mA. A doctor would once again prescribe a specific dose for the patient [28,29].

Finally, the number of sessions required to attain the best neurological and cognitive results is an important aspect of the treatment. According to Castillo-Saavedra et al. [30], the ideal number of sessions per week was five. Loo et al. [20] found similar outcomes with treatments lasting between two and eight weeks. After week six, however, no additional gains were detected. There was a potential of modest unfavorable effects on the patients if the number of sessions was surpassed. [31]. As a result, the platform must include a scheduling or control mechanism to ensure that the patient is protected according to the doctor's orders.

There have been randomized sham tDCS trials, in which the device implies to the patient that the system is giving the current to the patient, in order to verify any tDCS system is successful in treating a patient. In actuality, no current is applied; this is known as a sham or placebo tDCS study [32,33]. While such papers illustrate how to conduct sham tDCS trials, they don't go into detail about a device-specific way that would allow the hardware platform to automate the procedure by giving patients both false and genuine treatments. Previous research has only suggested a random crossover mechanism in the midst of a study by randomly assigning patients to sham or genuine tDCS treatments [32]. As an outcome, more research into automating the integration of placebo and genuine therapies into the hardware platform is required.

### 2.2.3. TRANSCRANIAL ALTERNATING CURRENT STIMULATION (TACS)

Transcranial Alternating Current Stimulation (tACS) is a cutting-edge non-invasive brain stimulation technique that holds promise for understanding and modulating brain activity [34]. By applying weak alternating electrical currents to the scalp, tACS aims to influence brain oscillations and neural synchronization at specific frequencies [35]. The alternating current induces changes in the excitability of neurons, leading to the entrainment of brainwave patterns associated with various cognitive functions. Unlike Transcranial Direct Current Stimulation (tDCS), which delivers a constant electrical current, tACS specifically targets the frequency of the brain's natural electrical rhythms, allowing researchers to fine-tune the effects and achieve more precise and frequency-specific brain modulation. This makes tACS a compelling tool for investigating the causal relationship between brain oscillations and cognitive processes and exploring its potential applications in enhancing cognition or treating neurological and psychiatric disorders.

### 2.2.4. TRANSCRANIAL RANDOM NOISE STIMULATION (TRNS)

Transcranial Random Noise Stimulation (tRNS) is another non-invasive brain stimulation technique that involves applying random electrical noise to the scalp to modulate brain activity [36]. Similar to Transcranial Alternating Current Stimulation (tACS), tRNS aims to influence neural excitability and brain oscillations. However, unlike tACS, which uses a specific alternating current frequency, tRNS delivers random electrical noise covering a broad frequency range [37]. The random noise in tRNS is thought to increase the overall neural excitability in the targeted brain regions, making neurons more responsive to incoming stimuli and potentially enhancing cortical plasticity. The mechanism of action is not fully understood, but it is believed that tRNS may cause a random firing of neurons, leading to a kind of "stochastic resonance" effect, where noise enhances the detection and transmission of weak signals in the brain.

One advantage of tRNS is that it does not require fine-tuning the stimulation frequency, making it a simpler and potentially more broadly applicable technique compared to tACS [38]. Additionally, tRNS may have advantages in certain situations where the optimal frequency for brain modulation is unknown or where multiple frequencies may be involved in a particular cognitive process. tRNS is being explored in research and clinical settings to study its effects on various cognitive functions, learning, memory, motor skills, and as a potential treatment option for neurological and psychiatric disorders. As with other brain stimulation techniques, tRNS also requires careful investigation and supervision to ensure its safety and efficacy in specific applications.

## 3. PROPOSED IOMT-BLOCKCHAIN BASED TDCS FRAMEWORK

Our suggested tDCS platform includes cloud communication as a crucial component to ensure both patient and physician privacy and confidentiality. Patients are treated remotely after getting the specialty gadget, and doctors deal with patients remotely via video conference, according to studies [39]. Despite the fact that this framework provided a way to provide a patient's needed dosage in accordance with existing tDCS safety rules and recommendations, [5, 26] it did not deliver a guideline for grasping real-time data about individual treatment parameters or details regarding patient's conditions. It also doesn't go over further safety features like the device's ability to administer the right amount to the patient or the doctor's ability to operate it remotely. This research presents a unique IoMT-blockchain-based architecture for bi-directional communication between a patient's tDCS device held remotely (such as at home) and a physician's software interface. Figure 2 demonstrates the treatment process in the proposed tDCS framework.

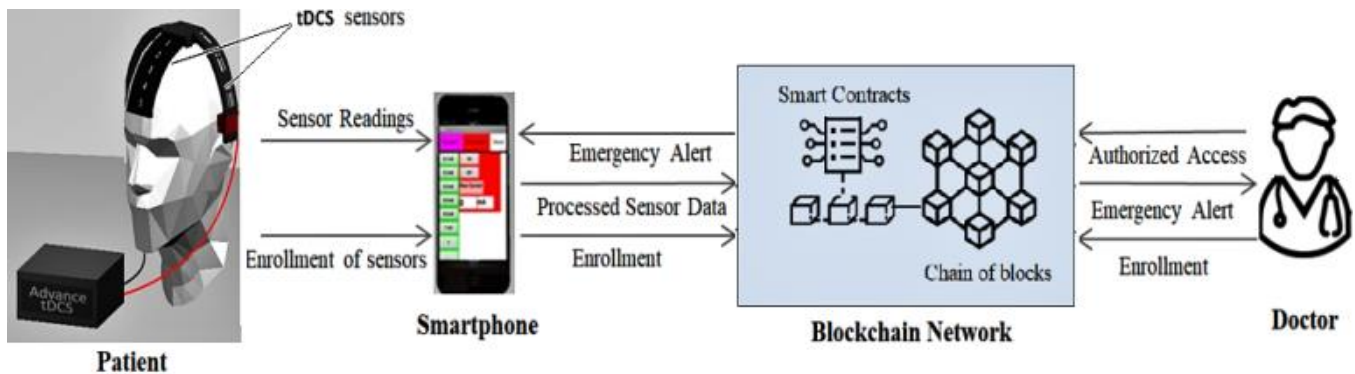

**Figure 2.** Proposed framework for IoMT-Blockchain based neuro-stimulation system.

Patients and doctors are the two sorts of users who connect to the system via their cell phones. The system contains all of the components of a distributed network, with the exception of a blockchain network at the network layer, which offers all of the distributed ledger technology's capabilities.

### A. COMPONENTS OF THE PROPOSED TDCS FRAMEWORK

The components of this proposed framework are briefly explained in this section.
**Patients**: Each patient will become a node in the network. To read health data, various IoT gadgets will be implanted on their bodies. The data from the sensors will be collected and processed by a mobile app on the patient's smartphone before being sent to the network. As a result, in our system, the mobile app may be thought of as a virtual patient.

**Doctors:** Doctors are registered individuals who are in charge of monitoring and treating the patients in the system.

**Blockchain Network**: The system will have a blockchain network that will connect all of the system's components. To be included in this network, all of the nodes in it must be confirmed. Because the data must be accessible to everybody, the network will be peer-to-peer, and the blockchain will be permissioned. The network is devoid of rogue nodes thanks to permissioned blockchain. Because there is no rogue node in the system, the PBFT consensus process should be used to ensure the legitimacy of each transaction in the network. Healthcare data will not be directly kept in the blockchain; rather, data storage and access will be recorded as blockchain transactions. The patients' processed data should be saved on a cloud server. To provide safe data access, all users should have digital signatures. For real-time monitoring, each transaction must be coupled with many smart contracts. They will be activated in response to data values and peer behavior.

**Cloud Storage:** Because monitoring data is acquired on a continual basis and must be retained in the system, the amount of healthcare data will grow with time. If the data is stored on a blockchain ledger, the devices at the user's end will require a large amount of storage. Furthermore, if we wish to store data in the blockchain, a node's disconnection may result in data loss. As a result, cloud storage may be used to store actual data, with the blockchain network storing the link to the data in the cloud server as part of the transaction.

### B. Operations in the Framework

The tDCS architecture we propose focuses on continuously reading health data, storing it, and notifying authorized people in the event of an emergency. As a result, we can highlight three key processes that must be completed within the constraints.

i. **Doctors' Registration and Assignment**: Patients and doctors will request registration providing all the information to the hospital. Then, the hospital authority will validate them using a defined smart contract in the system. Because of smart contracts, the system is responsive and act in real-time [40]. After being validated by the smart contracts, information of the patients and doctors will be added to the ledger of the blockchain. Lastly, hospital authorities will assign a doctor to a patient.

ii. **Data collection and storage on a continuous basis**: All health data is constantly gathered and kept in the system, and any odd occurrence must be reported to the appropriate parties. Raw sensor and device values will be delivered to the patient's smartphone or tablet through a mobile application. The raw data is then formatted and processed by the mobile application. The data is processed and then passed through a smart contract for additional analysis. Because the system requires real-time monitoring, smart contracts should be developed to detect any abnormalities in the patient's condition based on the monitoring data. The data from the smart contract should be kept in the cloud storage after it has been analyzed. Each information upload and data access event must be recorded in the ledger as a transaction. All monitoring data should be kept in the cloud (Figure 3) and adhere to the Health Insurance Portability and Accountability Act (HIPAA) procedure to prevent data mapping.

```
Patient name: Haydar Mahmud
Treatment lenght: 5 mins
Date:2021/9/21, Time:13-1-7,  Current:1,
Date:2021/9/21, Time:13-1-23, Current:1,
Date:2021/9/21, Time:13-1-26, Current:1,
Date:2021/9/21, Time:13-1-29, Current:1,
Date:2021/9/21, Time:13-1-31, Current:1,
Date:2021/9/21, Time:13-1-33, Current:1,
Date:2021/9/21, Time:13-1-36, Current:1,
Date:2021/9/21, Time:13-6-57, Current:1,
Date:2021/9/21, Time:13-7-30, Current:0,
Date:2021/9/21, Time:13-8-5,  Current:25,
```

**Figure 3.** tDCS Session data stored in cloud

iii. **Data Requests and Access**: Doctors may also view a patient's monitoring data, which is an important feature of the system. A request is issued to the system when a doctor wants to view a patient's data. If the patient wants the doctor to have access to his or her monitor data, he or she will provide the doctor his or her key in exchange for the request. When a doctor submits a request to the system, a smart contract is launched to verify the doctor's identity before granting access to the patient's data. After being authenticated in the system, the doctor may access the patient's data from the ledger.

## 4. Hardware Prototype of Proposed tDCS Framework

The system framework is depicted in Figure 4 as a block diagram. The essential component of a tDCS device is the circuit board. A processor ATMEGA328P handles this circuit board. Some current-regulating ICs make up the IC panel (LM334). The amount of current supplied by each IC is varied. On the basis of current requirements, the CPU adjusts the supply power to power up the ICs. The CPU is linked with a real-time clock and an SD card module. The stimulation period is counted in real time. The real-time data is stored in the cloud module. ATMEGA328P is also linked to a Bluetooth device. Bluetooth establishes a link between the mobile app and the circuit board. Anode refers to the positive side of an electrode. It connects to the IC panel and delivers the output to the brain's scalp. The cathode is the electrode that is linked to the ground and is the negative component of the electrodes. A power source is included with the board to provide the system with the necessary electricity.

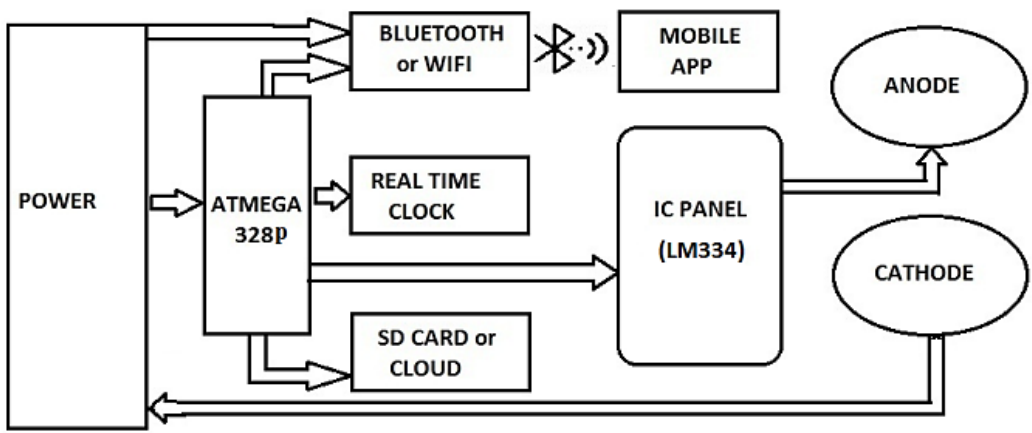

**Figure 4**. Block diagram of the system.

We utilized LM334 with a digital potentiometer to control a little amount of current (digi-pot). For different values of digi-pot, we receive varied amounts of current from this circuit (Figure 5). The main idea for obtaining the desired amount of current at the electrodes of this device is as follows.

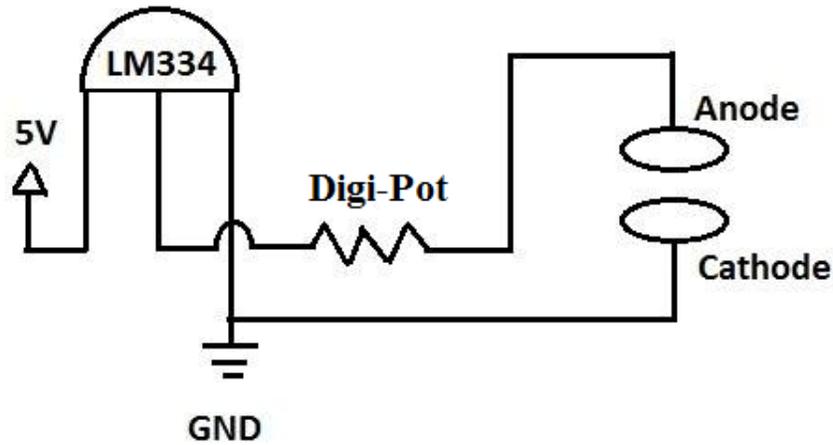

**Figure 5.** Current regulating circuit

One of the most crucial components of this tDCS gadget is the headband (Figure 6). The headband has a numerical basis that goes from (–V) to (+V). The base is made up of two hands that are numbered from (I) to (V) (III). These two hands are able to move in tandem with the base. Each hand has two square compartments in which electrodes are put. These square boxes may also be moved around freely, allowing electrodes to be put wherever on the user's head. Different disorders necessitate various electrode locations.

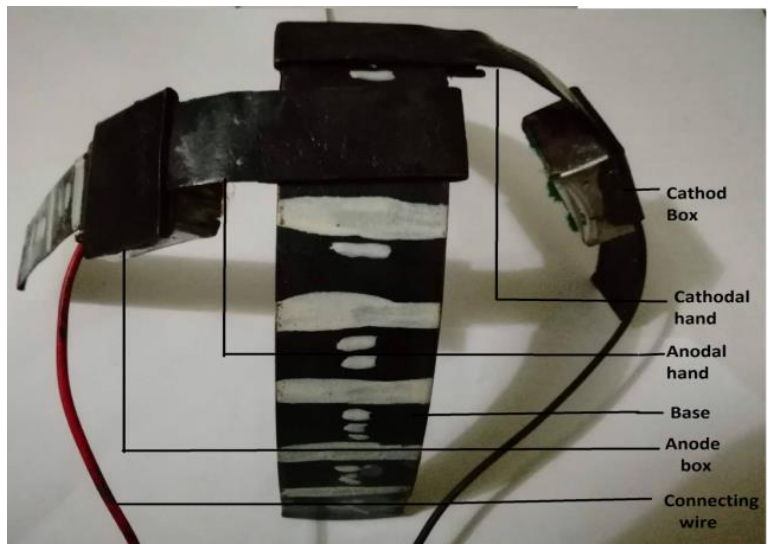

**Figure 6.** tDCS Headband

The circuit board can be viewed in Figure 7 to show all of the system's components. The board now has a real-time clock and cloud module, allowing the system to retain current passing data in real time and date. This information will be useful to both the user and the clinician for future treatment.

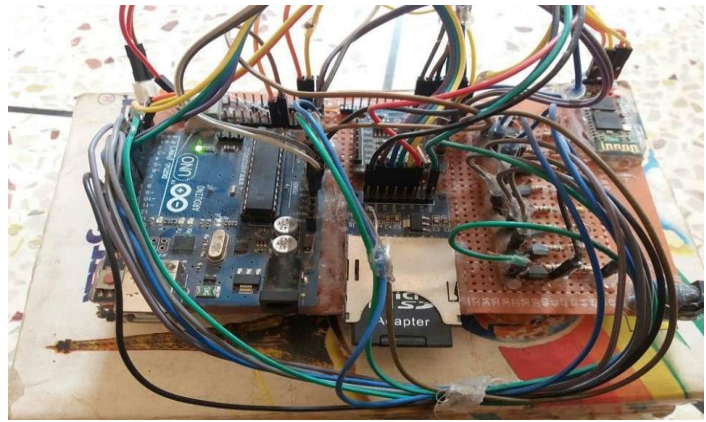
**Figure 7.** The circuit board of the device

To control this advanced tDCS gadget, we created an Android app (Figure 8). This software must use Bluetooth to communicate with the circuit board. From this app, we can provide commands for the needed current.

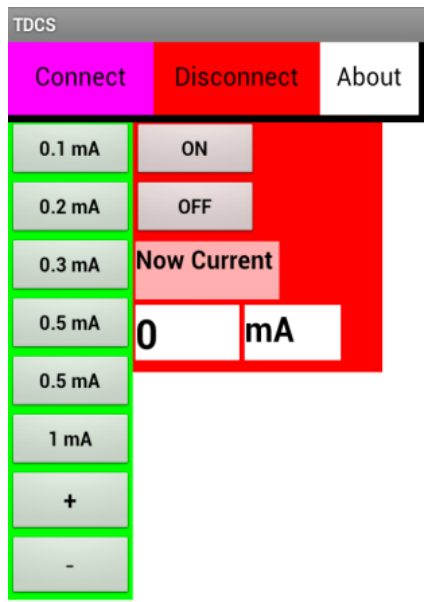
**Figure 8.** Mobile app interface.

## 5. CONCLUSION AND FUTURE DIRECTION

We intended to address a pressing issue that requires more attention in the field of biomedical engineering research. This research focuses on using blockchain and Internet of Medical Things (IoMT) technologies to provide efficient and secure remote patient monitoring (RPM) in tDCS devices. In the future, the prototype should be enhanced more so that it may be utilized in clinical trials and provide a more seamless experience for physicians. The IoMT-blockchain-based neurostimulation solutions may be used with machine learning technologies to automatically customize smart contracts for each patient. Such integration is something we'll work on in the future.